\documentclass[aps,reprint,twocolumn,superscriptaddress,nofootinbib,longbibliography,nobalancelastpage,prb]{revtex4-2}
\usepackage{graphicx}
\usepackage{float}
\usepackage[caption=false]{subfig}
\usepackage{amssymb}
\usepackage{amsmath}
\usepackage{amssymb}
\usepackage{amsfonts}
\usepackage{bm}
\usepackage{braket}
\usepackage[dvipsnames]{xcolor}
\usepackage[colorlinks=True,citecolor=blue,urlcolor=blue,linkcolor=red]{hyperref}
\usepackage{hypcap}

\DeclareMathOperator{\ceil}{ceil}
\DeclareMathOperator{\hc}{h.c.}

\newcommand{\hp}[1]{\ensuremath{H^{(#1)}}}

\begin{document}

\title{Stark Many-Body Localisation Under Periodic Driving} 

\begin{abstract}

We study stability of localisation under periodic driving in Stark many-body  systems. We find that localisation is stable except near special resonant frequencies, where resonances cause delocalisation. We provide approximate analytical arguments and numerical evidence in support of these results. This shows that disorder-free broken ergodicity is stable to driving, opening up the way to studying nonequilibrium driven physics in a novel setting.

\end{abstract}

\author{Christian Duffin}
\affiliation{Interdisciplinary Centre for Mathematical Modelling and Department of Mathematical Sciences, Loughborough University, Loughborough, Leicestershire LE11 3TU, United Kingdom}

\author{Aydin Deger}
\affiliation{Interdisciplinary Centre for Mathematical Modelling and Department of Mathematical Sciences, Loughborough University, Loughborough, Leicestershire LE11 3TU, United Kingdom}
\affiliation{Department of Physics and Astronomy, University College London, London WC1E 6BT, United Kingdom}

\author{Achilleas Lazarides}
\affiliation{Interdisciplinary Centre for Mathematical Modelling and Department of Mathematical Sciences, Loughborough University, Loughborough, Leicestershire LE11 3TU, United Kingdom}

\maketitle

\section{Introduction}
Generic ergodic quantum systems are expected to thermalise according to the predictions of statistical mechanics. Such ergodic systems are understood to be in agreement with the eigenstate thermalisation hypothesis (ETH) \cite{Deutsch:1991ju,Srednicki:1994dl,DAlessio2015}, which states that the eigenstate expectation values of any local observable form a smooth function of energy. Any deviation from this expectation is, by default, interesting, as it leads to the unusual deviations from statistical mechanics. More practically, non-adiabatic manipulations of ergodic systems leads to runaway entropic increase and heat death~\cite{Lazarides:2015jd}; therefore, creating non-equilibrium phases such as time crystals~\cite{Khemani2016,Else:2016ue,Yao2017} \emph{requires} breaking ergodicity, as usually these rely on periodic driving.

It is therefore interesting to study systems that exhibit non-ergodic behaviour. Ergodicity can be broken in various ways including glassiness, prethermalisation~\cite{DAlessio2015}, or kinetic constraints~\cite{Lan2018,Roy2019a,Deger22a,Deger22b}. More recently there has been considerable attention devoted to disorder-induced many-body localisation (MBL) \cite{Basko:2006hh,Oganesyan:2007ex,Znidaric:2008ux,Pal:2010gr}. In Anderson's original work, it was shown that transport in a tight-binding model for a single particle is suppressed by inherent disorder \cite{Anderson:1958fz}. MBL extends this phenomenon to the case of interacting systems. Similarly, it has been long known that transport can instead be suppressed by the presence of a tilted potential, leading to so-called Wannier-Stark localisation \cite{wannier1962dynamics}. Here, the single-particle eigenstates become localised in space and equidistant in energy by a spacing that corresponds to the gradient of the tilt. Only much more recently was this shown to also persist in the presence of interactions, giving what is known as Stark MBL \cite{schulz2019stark,Morong2021,taylor2020experimental,guo2021stark,zhang2021,doggen2021stark}.

Periodic driving of a noninteracting Wannier-Stark system has been studied thoroughly. It was found that driving at a resonant frequency determined by the tilt angle results in delocalisation for arbitrarily weak driving \cite{Holthaus1996,Shon1992}. Here, we investigate the analogous situation in an interacting system, finding that delocalisation does still occur near certain resonant frequencies, but only above a finite driving strength. We use approximate analytical arguments to explain how this driven system delocalises in the non-interacting limit, and extend this to the interacting case using exact numerical methods. This therefore establishes that the situation is analogous to that for disordered MBL systems~\cite{Lazarides:2015jd,Ponte:2015dc}. Previous work on driving Stark systems has focussed on the high frequency regime~\cite{Bhakuni2020}.

\section{Model and analytical arguments}

\subsection{Hamiltonian}
We will focus on the time-periodic Hamiltonian
\begin{equation}
    \label{eq:drive-clean}
        H(t)=
        \begin{cases}
        H_0, &  0 < \mod(t, T)\leq T_0 \\
        H_1, &  T_0 < \mod(t,T)\leq T
    \end{cases}
\end{equation}
with period $T=T_0+T_1$ and 
\begin{equation*}
    H_0=-\gamma\sum_j^Ljn_j+\sum_{j}^{L}h_{j}n_j + V\sum_j^{L-1}n_jn_{j+1} 
\end{equation*}
diagonal in the basis of product (Fock) states, while 
\begin{equation*}
    H_1= -J\sum_j^{L-1}(b_jb_{j+1}^{\dag}+\hc).
\end{equation*}
Here the $b_j$ are hard-core bosonic operators and our system lives on a 1D lattice of size $L$. The parameters $J$ and $V$ denote the amplitudes of nearest-neighbour hopping and interactions, respectively, while $\gamma$ denotes the gradient of the tilted potential. The times $T_0$ and $T_1$ correspond to the diagonal $H_0$ and off-diagonal $H_1$ respectively, such that $T_1$ controls the strength of the driving and the system is trivially localised in the \(T_1=0\) limit.We choose this model as it has been well established for studying driven systems \cite{Ponte:2015dc}.

\begin{figure}[t]
    \centering
    \includegraphics[width=1\linewidth]{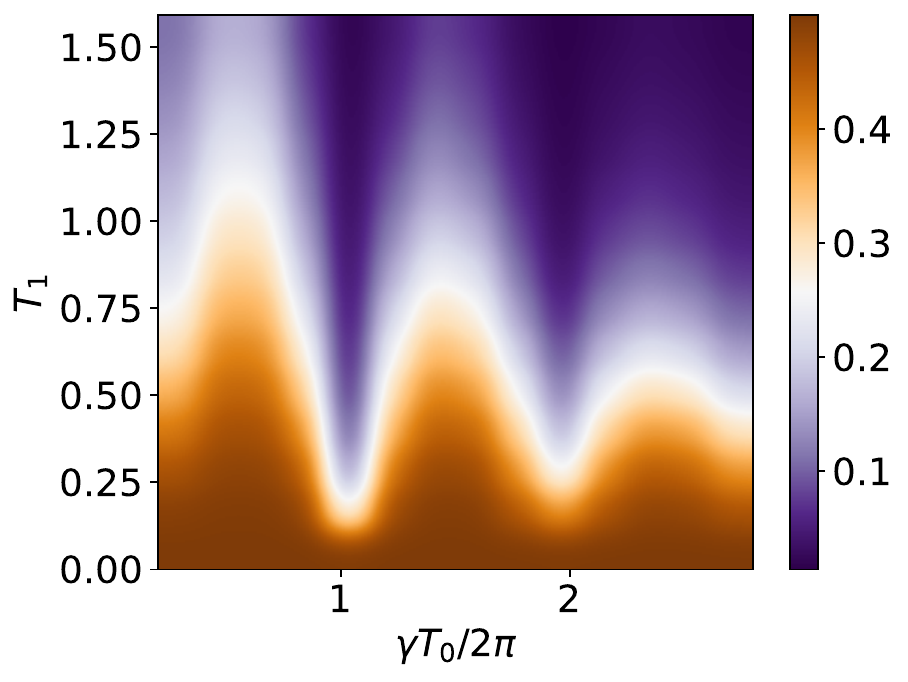}
    \caption{Phase diagram of a driven Stark MBL system for the unkicked (diagonal) period \(T_0\) vs kicked (off-diagonal) period \(T_1\). The metric used is the standard deviation \(\sigma\) of the probability distribution for the EEVs \(\langle n_6\rangle\), where \(n_6\) is the number operator for a particle on site 6 of a 1D lattice. The ergodic regime is represented by the purple region \(\sigma\approx0\), corresponding to all EEVs equaling 0.5. The MBL regime is represented by the orange region \(\sigma\approx0.5\), corresponding to all EEVs equaling 0 or 1. The valleys in the MBL regions at integer multiples of \(\gamma T_0/2\pi\) indicate resonant points, where driving occurs at a frequency that eliminates energy separations caused by the tilted potential with gradient \(\gamma\). Parameters used are \(L=14,\,J=V=0.5,\,\gamma=5,\,w=0.05\), and 14 data points are used for both axes.}
    \label{fig/EEV_map}
\end{figure}

In the limit $h_j=0$, the static system $H_0+H_1$ is not localised~\cite{kloss_absence_2023}, suggesting that the Floquet system described by Eq.~\ref{eq:drive-clean} will also be delocalised. We therefore add a small disorder term where the \(h_j\) are drawn from a zero-mean normal distribution with variance \(w\). To ensure that the disorder itself does not cause MBL, we set  $w\ll V$. This is enough for the static system to be localised~\cite{kloss_absence_2023}. Alternatively, one could also achieve this by introducing a curvature term to the tilted potential as a function of system size, however this brings with it the possible disadvantage of additional finite-size effects.~\footnote{Alternatively, in the limit of $T_1=h_j=V=0$, conservation of dipole moment means that all Fock states with the same dipole moment (centre of mass) will be degenerate. If then $T_1>0$, these states will strongly hybridise, delocalising the system. Adding weak disorder lifts the degeneracy.} 

Throughout this work, the parameters used will be \(J=V=T_1=0.5,\,\gamma=5,\,w=0.05\) (See Appendix A for motivation). Finally, we work at half-filling and with open boundary conditions. 

The Floquet operator for this Hamiltonian, propagating the system over a period, is then
\begin{equation}
    \mathcal{F}=e^{-iH_0T_0}e^{-iH_1T_1}.
\end{equation}

\subsection{Mapping to RZS}

One can obtain analytical insight into the behaviour of this system by recasting the problem in the so-called Repeated Zone Scheme (RZS)~\cite{Shirley:1965cy}. To facilitate this, we recast the Hamiltonian of Eq.~\ref{eq:drive-clean} as
\begin{equation}
    H(t)=\bar{H}+\delta H(t),
    \label{eq:H-clean-recast}
\end{equation}
where the time-independent part \(\bar{H}\) is the time average of \(H(t)\) over a single period while \(\delta H(t)\) is the rest. We have
\begin{equation}
        \begin{split}
        \bar{H} &= H_0\frac{T_0}{T}+H_1\frac{T_1}{T}\\
        &=\sum_j
            \left[
                \tilde{J}(b_jb_{j+1}^{\dag}+\hc)
                +\tilde{V}n_jn_{j+1}
                +\left(\tilde{\gamma}j+\tilde{h}_j\right)n_j
            \right]
    \end{split}
    \label{eq:H-0-effective}
\end{equation}
with \(\tilde{J}=JT_1/T,\,\tilde{V}=VT_0/T,\,\tilde{\gamma}=\gamma T_0/T\), and \(\tilde{h}_j=h_jT_0/T\) and
\begin{align}
    \delta H(t)= 
        \begin{cases}
        \delta H_1=\bar{H}\frac{T}{T_0}-H_1\frac{T_1}{T_0},  \mod(t,T)\leq T_0 &\\
        \delta H_2=\bar{H}\frac{T}{T_1}-H_0\frac{T_0}{T_1}, ~T_0<\mod(t,T)\leq T& .
        \end{cases} 
\end{align}

One can eliminate the time dependence at the price of working with infinitely many coupled undriven systems~\cite{Shirley:1965cy}. Expanding in Fourier components, one finds that our system corresponds to a time-independent system on an infinite ladder, the $n^\mathrm{th}$ rung of which hosts a copy of the static part of the Hamiltonian $\overline{H}$ shifted in energy by $n\omega$. The rungs are then coupled by the Fourier components of the driving Hamiltonian, $\delta H(t)$ which cause hopping between the rungs. 
Denoting by $\ket{\alpha}$ a basis state of the Hilbert space of $\overline{H}$, the time-independent system can be written in terms of the states $\ket{\alpha,n}=\ket{\alpha}\otimes\ket{n}$ localised in rung $n$ and the physical basis state $\alpha$ in the form
\begin{equation}
    \bra{\alpha,n}H_F\ket{\beta,m} = \hp{n-m}_{\alpha,\beta}-\omega n\delta_{n,m}\delta_{\alpha,\beta}.
    \label{eq:H-shirley}
\end{equation}
Here, $\hp{p}_{\alpha,\beta}$ are the matrix elements $\bra{\alpha}\hp{p}\ket{\beta}$ of the $p^{th}$ Fourier component of the physical Hamiltonian, $\hp{p}$. 
\newcommand{\deltaH}[1]{\ensuremath{\delta H^{(#1)}}}
\newcommand{\Have}{\ensuremath{\overline{H}}}
Writing $H^{(0)}=\Have$ and $H^{(p)}=\deltaH{p}$ for $p\neq 0$ (since these are the Fourier components of $\delta H$ of Eq.~\ref{eq:H-clean-recast}), $H_F$ can be represented as follows:
\begin{equation}
  H_F=\left(
    \begin{array}{c|c|c|c|c|c|c}
        \ddots  & \vdots  & \vdots  & \vdots & \vdots & \vdots & \reflectbox{$\ddots$} \\
        \hline    
        \cdots & \Have -2\omega   &\deltaH{1}     & \deltaH{2}    &\deltaH{3} & \deltaH{4} & \cdots \\
        \hline  
        \cdots & \deltaH{1}       &\Have-\omega   & \deltaH{1}    &\deltaH{2}  & \deltaH{3} &\cdots \\
        \hline  
        \cdots &\deltaH{2} &  \deltaH{1}      & \Have         &  \deltaH{1}& \deltaH{2} & \cdots \\
        \hline  
        \cdots &\deltaH{3}&  \deltaH{2}      & \deltaH{1}    & \Have+\omega & \deltaH{1} & \cdots \\
        \hline
        \cdots &\deltaH{4}&  \deltaH{3}      & \deltaH{2}    & \deltaH{1} &\Have+2\omega  & \cdots \\
        \hline  
        \reflectbox{$\ddots$} & \vdots & \vdots & \vdots & \vdots & \vdots & \ddots
    \end{array}
           \right)
           \label{eq:HF-visual}
\end{equation}
Physically, the hopping between different rungs corresponds to the system gaining or losing energy in integer multiples of $\omega$, and is mediated by the components
\begin{equation}
    \deltaH{n}=\frac{\delta H_0}{2\pi i n}\left(\mathrm{e}^{2\pi inT_0/T}-1\right)
        +
    \frac{\delta H_1}{2\pi i n}\left(\mathrm{e}^{2\pi in}-\mathrm{e}^{2\pi inT_1/T}\right).
    \label{eq:fourier-components}
\end{equation}
We note for later reference that the components decay algebraically with $n$.

Solutions of the time-dependent Schr\"odinger equation $\left(i\partial_t-H(t)\right)\ket{\phi(t)}=0$ can be written 
as $\ket{\phi_a(t)}=e^{-i\Omega_a t}\sum_{n}\ket{\varphi_a^{(n)}}e^{-in\omega t}$, 
where 
$\ket{\varphi_a}=\sum_{n=-\infty}^\infty\ket{\varphi_a^{\scriptscriptstyle{(n)}}}\otimes\ket{n}$ 
is an eigenstate of the time-independent Hamiltonian \eqref{eq:HF-visual} with eigenvalue $\Omega_a$. Here $\ket{n}$ indexes the rung, so that $\ket{\varphi_a^{\scriptscriptstyle{(n)}}}$ is the part of the state living in a single rung. For stroboscopic dynamics, $\ket{\phi_a(pT)}=e^{-i\Omega_a pT}\sum_{n}\ket{\varphi_a^{(n)}}$ (for integer $p$) so that one sums up the across all rungs.

What does this picture say about the system of Eq.~\ref{eq:H-clean-recast}? In the limit we are working in ($h_j=V=0$) and $\delta H(t)=0$, the $\ket{\varphi_a^{(n)}}$ are localised and independent of $n$; they are the eigenstates $\ket{\alpha}$ of $\Have$, which are many-body Stark localised states with eigenvalues $q_\alpha\tilde\gamma$ where the $q_\alpha$ are integers.\footnote{This is because the single-particle states have energies $m\tilde\gamma$ with $m$ integer, and a many-body state is simply a Slater determinant of such states thus having an energy that is a sum of these.} 

To delocalise the system the driving must couple $\ket{\varphi_a^{(n)}}$ with different $a$ in different rungs $n$, so that the stroboscopic dynamics of the resulting state will be delocalised. Perturbatively in $\delta H$, there are two leading-order ways this can happen. 

The first is that two eigenstates $\ket{\alpha}$ and $\ket{\beta}$ in neighbouring rungs are coupled by the matrix element $\bra{\alpha}\deltaH{1}\ket{\beta}$; the energy difference is $\overline{\epsilon}_\alpha-\overline{\epsilon}_\beta=m\tilde\gamma$ with integer $m$; since they are on neighbouring rungs, the condition on $\omega$ for this is $\overline{\epsilon}_\alpha-\overline{\epsilon}_\beta=m\tilde\gamma=\omega$ from which we find $\frac{\gamma T_0}{2\pi}=\frac{1}{m}.$

Now the matrix element $\bra{\alpha}\deltaH{1}\ket{\beta}$ is exponentially suppressed whenever $\ket{\alpha}$ and $\ket{\beta}$ differ by moving a particle by more than one site (because $\deltaH{1}$ has only nearest-neighbour hopping and the $\ket{\alpha}$ are localised). Thus $m=1$ is dominant and we expect delocalisation to be clearly visible at $\frac{\gamma T_0}{2\pi}=1.$

The second is that two eigenstates $\ket{\alpha}$ and $\ket{\beta}$ in rungs $n$ and $n+p$, with $p>1$, respectively are coupled by $\bra{\alpha}\deltaH{p}\ket{\beta}$. The resonance condition is then $\overline{\epsilon}_\alpha-\overline{\epsilon}_\beta=m\tilde\gamma=p\omega$ leading to $\frac{\gamma T_0}{2\pi}=\frac{p}{m}$. Like before, $m>1$ is exponentially suppressed, while the matrix element $\bra{\alpha}\deltaH{p}\ket{\beta}$ decays linearly with $p$; since $p>1$, $p=2$ dominates and we expect strong signs of delocalisation at $\frac{\gamma T_0}{2\pi}=2,$ with weaker delocalisation for increasing $p$.

Thus there are resonances at
\begin{equation}
    \gamma T_0/2\pi=p/m
    \label{eq:resonances}
\end{equation}
for integer $p,m$, with $p=m=1$ the dominant one, followed by $p=2,m=1$, then $p=3,m=1$ and so on. The ones for $m>1$ are strongly suppressed because of the locality of $\delta H(t)$ and will not be visible in our numerics. These conclusions are in broad agreement with the results of Holthaus and Hone on single-particle systems~\cite{Holthaus1996}.

Away from the limit $h_j = V = 0$, this picture is no longer exact. We therefore use numerics to study to what extent the conclusions survive away from this limit. To check whether the system is localised we implement two numerical probes of localisation, and find that they do indeed confirm our conclusions. Fig. \ref{fig/EEV_map} shows a summary of the numerical calculations that we now describe. 

\section{Eigenstate properties}
A localised system violates the ETH, while delocalisation via periodic driving is expected to coincide with its reestablishment~\cite{Lazarides:2015jd,Ponte:2015dc}. We therefore probe directly ETH by checking whether the eigenstate expectation values (EEVs) form a smooth function of the eigenenergies or not. Selecting our local observable to be the number density of a particle on some fixed site $j$, we calculate the EEV for each eigenstate \(|\epsilon_\alpha\rangle\) as \(\langle n_i\rangle=\langle\epsilon_\alpha|n_i|\epsilon_\alpha\rangle\), and plot these versus the quasienergies $\epsilon_\alpha=-\left(\ln\Theta_\alpha\right)/iT$
where \(\Theta_\alpha\) are the eigenvalues of the propagator $\mathcal{F}$ corresponding to $\ket{\epsilon_\alpha}$.

\begin{figure}[t]
    \centering
    \includegraphics[width=\columnwidth]{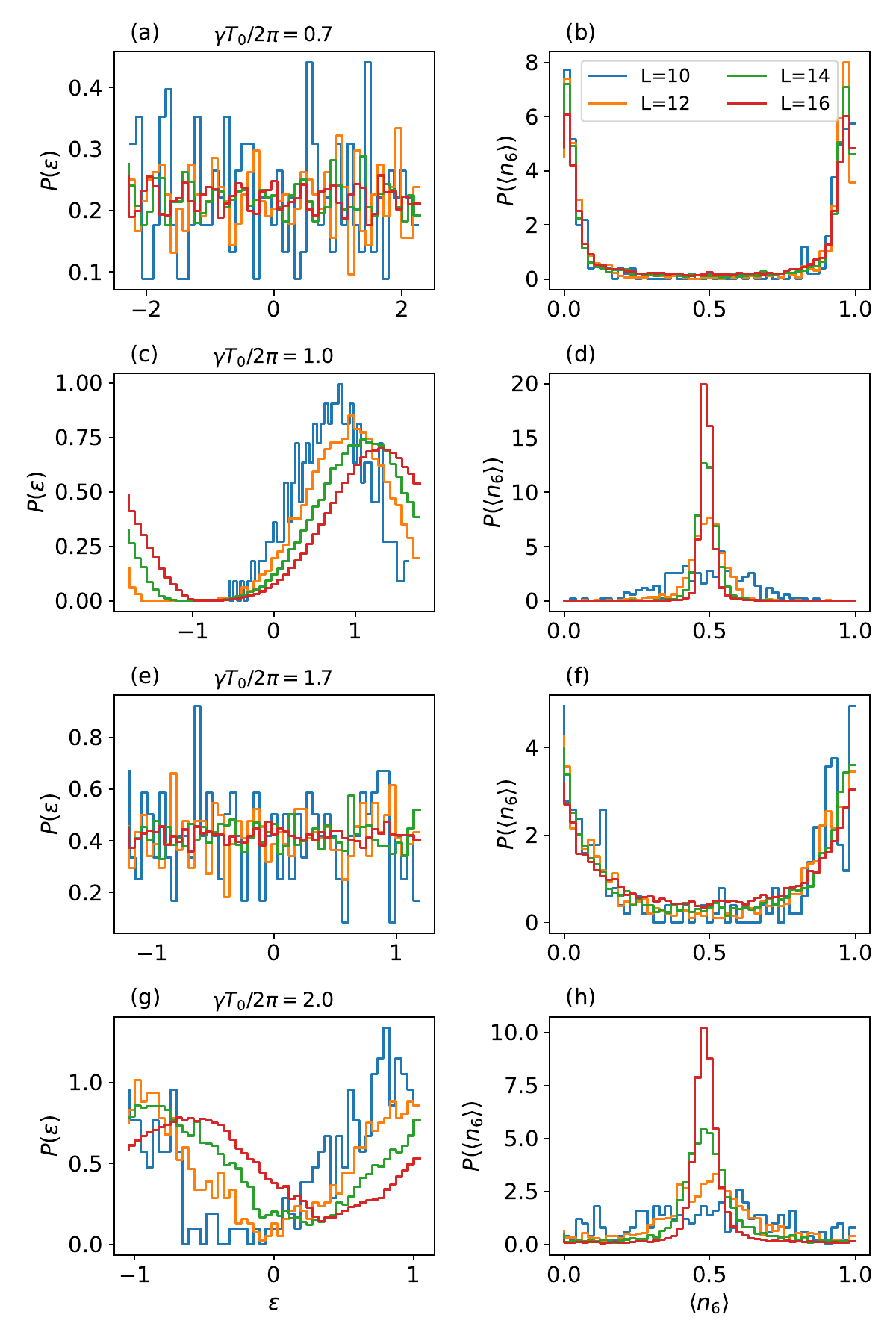}
    \caption{Left column: Density of states as a function of quasienergy \(\epsilon\). Right column: Probability distributions of eigenstate expectation values of the particle number density on site $j=6$, \(\langle n_6\rangle\) for a periodically driven Stark MBL system, at various system sizes \(L\) and unkicked periods \(T_0\). The second and fourth rows correspond to resonant driving at the resonant points \(n=1.0\) and \(n=2.0\) respectively, where the strongly peaked behaviour around \(\langle n_6\rangle=0.5\) in figures (d) and (h) indicates a strong lack of energy dependence between the EEVs, which corresponds to the infinite temperature ensemble that is associated with systems exhibiting Floquet ETH. The remaining figures show off-resonant driving, where the EEVs are strongly peaked around \(\langle n_6\rangle=0\) and \(\langle n_6\rangle=1.0\). This indicates violation of the ETH and thus the presence of MBL. Parameters used are \(J=V=T_1=0.5,\,\gamma=5,\,w=0.05\).}
    \label{fig/eev_hist}
\end{figure}

 In a maximally localised system half of the EEVs will equal 1 and the rest 0, while in an infinite-temperature system, as is expected for a driven, delocalised, ergodic system~\cite{Lazarides:2014ie}, they will all be equal to each other and to the infinite-temperature result $\mathrm{tr} \left(n_j\right)=1/2$~\cite{Lazarides:2015jd}.

In Figure \ref{fig/eev_hist} we plot both the density of states (DOS) as a function of quasienergy (left column) and the probability densities of the EEVs (right column) for various system sizes, both on resonance and off resonance, with the resonant points as determined earlier. 

In the off-resonant plots, the EEVs are evenly distributed between both \(\langle n_6\rangle=0\) and \(\langle n_6\rangle=1\) as anticipated for the localised case. On the other hand, for resonant driving the EEVs become sharply peaked at \(\langle n_6\rangle=0.5\), consistent with Floquet ETH. Crucially, the standard deviation decreases with system size, indicating that the system is indeed delocalised in the thermodynamic limit.

To further assess whether our results are reliable, we turn to the DOS plots. In the thermodynamic limit, we expect the DOS to be uniform due to the folding of the quasienergies and the mixing of the states caused by the driving. Indeed, in the off-resonant cases the DOS fluctuates around a constant value, with the amplitude of the fluctuations decreasing with system size. This shows that our driving protocol is able to mix the bands that are present in the undriven $T_1=0$ case (discussed in Appendix B).

In contrast, in the resonant cases the DOS instead shows strong quasienergy dependence. This suggests that our driving is not strong enough, or not low-frequency enough (compared to the static system's bandwidth), to mix the bands. From this viewpoint, increasing the system size will eventually result in mixing, which suggests that we are far from the thermodynamic limit. Nevertheless, this is not an issue: The EEVs already indicate delocalisation, which is favoured by better mixing of the bands. Thus, if anything the non-uniform DOS shows that our results \emph{underestimate} delocalisation, which is fine given that our result is that delocalisation occurs at all. Finally, the DOS becomes more evenly distributed with system size, which further supports the claim that this is a finite-size effect. 

\begin{figure}[t]
    \centering
    \includegraphics[width=1.05\columnwidth]{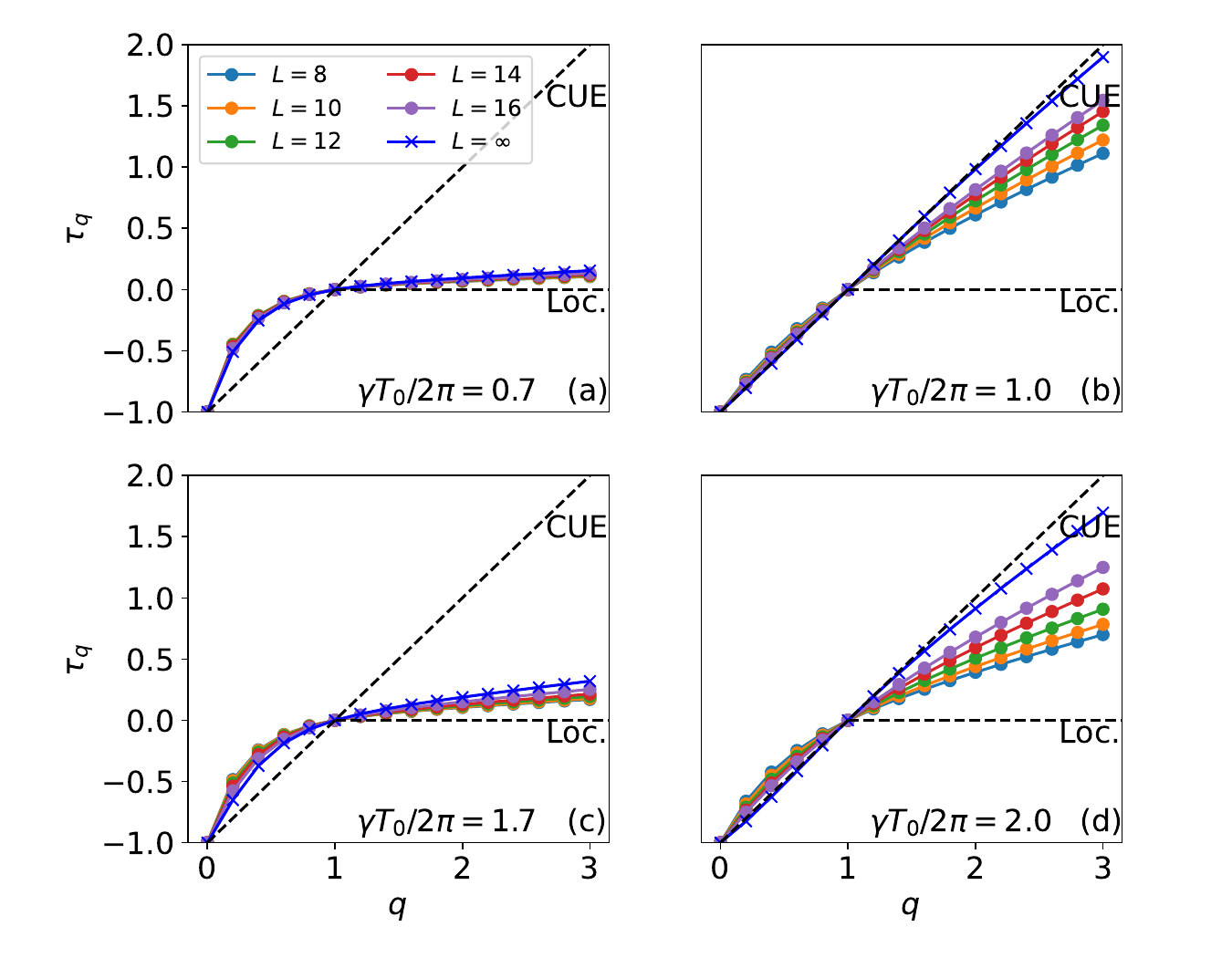}
    \caption{Averaged scaling exponent \(\tau_q\) at finite sizes and infinite size for the driven Stark MBL system under periodic driving. Averaging takes place over all eigenvectors and \(m\) disorder realisations, where \(m=\ceil(10000/\mathcal{D})\). Figures (a) and (c) correspond to off-resonant driving, which shows the expected multifractal behaviour for MBL eigenstates, specifically that the \(\tau_q\) sit above the thermal prediction at \(0<q<1\) and below it at \(q>1\). Figures (b) and (d) shows resonant driving at two different resonant points, which shows agreement with the thermal prediction up to values above \(q=1\), indicating that the eigenstates lie in the ETH regime. Parameters used are \(J=V=T_1=0.5,\,\gamma=5,\,w=0.05\).}
    \label{fig/tau}
\end{figure}

\section{Multifractality}

Another signature of broken ergodicity is accessed by studying the structure  of the eigenstates of $\mathcal{F}$. In an ergodic Floquet system, an eigenstate is simply a uniform superposition of all Fock states~\cite{Lazarides:2014ie}. Conversely, in MBL systems the eigenstates are multifractal in Fock space~\cite{DeLuca2013}. This multifractal property can be studied by invoking a metric called generalised inverse participation ratios (IPRs) \cite{Luitz2020}. This is defined by
\begin{equation}
    I_q=\sum_j|\langle\epsilon|j\rangle|^{2q}.
\end{equation}
At \(q=2\) this is the usual IPR. This depends on the Hilbert space dimension $\mathcal{D}$ via $I_q\sim\mathcal{D}^{-\tau_q}$ with $\tau_q=-\frac{\ln I_q}{\ln\mathcal{D}}$ the scaling exponent.  The \(\tau_q\) are related to the multifractal dimension \(D_q\), via
\(D_q=\frac{\tau_q}{q-1}\). 

In a maximally localised system (in Fock space), $D_q=0$ for $q>0$, which translates to $\tau_q=0$. In a completely delocalised system, $D_q=1$, which leads to the random matrix theory (RMT) prediction $\tau_q=q-1$. The general behaviour for disordered interacting systems is intermediate~\cite{Luitz:2015iv,Luitz2020,serbyn2017thouless}: close to the transition point between MBL and ergodic phase, the eigenstates are  multifractal in both regimes.

In Fig.~\ref{fig/tau} we show the \(\tau_q\) for a given \(q\) at several size, as well as our thermodynamic limit prediction. This was obtained by plotting two finite size values against \(1/\ln\mathcal{D}\) and extrapolating to \(1/\ln\mathcal{D}=0\) \cite{Luitz2020}. In these plots we average over all eigenvectors, as well as a number $m$ of disorder realisations; we select $m=\ceil(10000/\mathcal{D})$.

On resonance (panels (b) and (d)), our extrapolated result for the thermodynamic limit agrees  with the thermal prediction up to moments \(q*>1\). This indicates that the system is in the ETH phase, with a residual multifractality confined to high order moments. Off resonance, the \(\tau_q\) sit above the thermal prediction in the interval \(0<q<1\), and below it for all \(q>1\), thus exhibiting behaviour similar to that observed in MBL systems.

\section{Conclusions}

We have showed that Stark-MBL systems may remain localised under periodic driving. Such systems are clean and translationally-invariant, in contrast to the usual disordered-MBL systems, which also have this property. We expect that phenomena usually studied in disordered cases should have analogues for Stark systems~\cite{bar_lev_discrete_2024}. There also remain several open questions: How quickly does delocalisation set in at resonances, and how does it depend on the detuning? Similarly, how slow is heating in the absence of a non-linear component to the potential, when the static system is known to be delocalised, but correlations spread very slowly~\cite{kloss_absence_2023}? 

\acknowledgements{
    A.D. and A.L. acknowledge support from EPSRC Grant No. EP/V012177/1. 
}


\begin{thebibliography}{35}%
\makeatletter
\providecommand \@ifxundefined [1]{%
 \@ifx{#1\undefined}
}%
\providecommand \@ifnum [1]{%
 \ifnum #1\expandafter \@firstoftwo
 \else \expandafter \@secondoftwo
 \fi
}%
\providecommand \@ifx [1]{%
 \ifx #1\expandafter \@firstoftwo
 \else \expandafter \@secondoftwo
 \fi
}%
\providecommand \natexlab [1]{#1}%
\providecommand \enquote  [1]{``#1''}%
\providecommand \bibnamefont  [1]{#1}%
\providecommand \bibfnamefont [1]{#1}%
\providecommand \citenamefont [1]{#1}%
\providecommand \href@noop [0]{\@secondoftwo}%
\providecommand \href [0]{\begingroup \@sanitize@url \@href}%
\providecommand \@href[1]{\@@startlink{#1}\@@href}%
\providecommand \@@href[1]{\endgroup#1\@@endlink}%
\providecommand \@sanitize@url [0]{\catcode `\\12\catcode `\$12\catcode
  `\&12\catcode `\#12\catcode `\^12\catcode `\_12\catcode `\%12\relax}%
\providecommand \@@startlink[1]{}%
\providecommand \@@endlink[0]{}%
\providecommand \url  [0]{\begingroup\@sanitize@url \@url }%
\providecommand \@url [1]{\endgroup\@href {#1}{\urlprefix }}%
\providecommand \urlprefix  [0]{URL }%
\providecommand \Eprint [0]{\href }%
\providecommand \doibase [0]{https://doi.org/}%
\providecommand \selectlanguage [0]{\@gobble}%
\providecommand \bibinfo  [0]{\@secondoftwo}%
\providecommand \bibfield  [0]{\@secondoftwo}%
\providecommand \translation [1]{[#1]}%
\providecommand \BibitemOpen [0]{}%
\providecommand \bibitemStop [0]{}%
\providecommand \bibitemNoStop [0]{.\EOS\space}%
\providecommand \EOS [0]{\spacefactor3000\relax}%
\providecommand \BibitemShut  [1]{\csname bibitem#1\endcsname}%
\let\auto@bib@innerbib\@empty
\bibitem [{\citenamefont {Deutsch}(1991)}]{Deutsch:1991ju}%
  \BibitemOpen
  \bibfield  {author} {\bibinfo {author} {\bibfnamefont {J.~M.}\ \bibnamefont
  {Deutsch}},\ }\bibfield  {title} {\bibinfo {title} {Quantum statistical
  mechanics in a closed system},\ }\href@noop {} {\bibfield  {journal}
  {\bibinfo  {journal} {Phys. Rev. A}\ }\textbf {\bibinfo {volume} {43}},\
  \bibinfo {pages} {2046} (\bibinfo {year} {1991})}\BibitemShut {NoStop}%
\bibitem [{\citenamefont {Srednicki}(1994)}]{Srednicki:1994dl}%
  \BibitemOpen
  \bibfield  {author} {\bibinfo {author} {\bibfnamefont {M.}~\bibnamefont
  {Srednicki}},\ }\bibfield  {title} {\bibinfo {title} {Chaos and quantum
  thermalization},\ }\href@noop {} {\bibfield  {journal} {\bibinfo  {journal}
  {Phys. Rev. E}\ }\textbf {\bibinfo {volume} {50}},\ \bibinfo {pages} {888}
  (\bibinfo {year} {1994})}\BibitemShut {NoStop}%
\bibitem [{\citenamefont {Luca~D'Alessio}\ and\ \citenamefont
  {Rigol}(2016)}]{DAlessio2015}%
  \BibitemOpen
  \bibfield  {author} {\bibinfo {author} {\bibfnamefont {A.~P.}\ \bibnamefont
  {Luca~D'Alessio}, \bibfnamefont {Yariv~Kafri}}\ and\ \bibinfo {author}
  {\bibfnamefont {M.}~\bibnamefont {Rigol}},\ }\bibfield  {title} {\bibinfo
  {title} {From quantum chaos and eigenstate thermalization to statistical
  mechanics and thermodynamics},\ }\href@noop {} {\bibfield  {journal}
  {\bibinfo  {journal} {Advances in Physics}\ }\textbf {\bibinfo {volume}
  {65}},\ \bibinfo {pages} {239} (\bibinfo {year} {2016})},\ \Eprint
  {https://arxiv.org/abs/https://doi.org/10.1080/00018732.2016.1198134}
  {https://doi.org/10.1080/00018732.2016.1198134} \BibitemShut {NoStop}%
\bibitem [{\citenamefont {Lazarides}\ \emph {et~al.}(2015)\citenamefont
  {Lazarides}, \citenamefont {Das},\ and\ \citenamefont
  {Moessner}}]{Lazarides:2015jd}%
  \BibitemOpen
  \bibfield  {author} {\bibinfo {author} {\bibfnamefont {A.}~\bibnamefont
  {Lazarides}}, \bibinfo {author} {\bibfnamefont {A.}~\bibnamefont {Das}},\
  and\ \bibinfo {author} {\bibfnamefont {R.}~\bibnamefont {Moessner}},\
  }\bibfield  {title} {\bibinfo {title} {Fate of many-body localization under
  periodic driving},\ }\href@noop {} {\bibfield  {journal} {\bibinfo  {journal}
  {Phys. Rev. Lett.}\ }\textbf {\bibinfo {volume} {115}},\ \bibinfo {pages}
  {30402} (\bibinfo {year} {2015})}\BibitemShut {NoStop}%
\bibitem [{\citenamefont {Khemani}\ \emph {et~al.}(2016)\citenamefont
  {Khemani}, \citenamefont {Lazarides}, \citenamefont {Moessner},\ and\
  \citenamefont {Sondhi}}]{Khemani2016}%
  \BibitemOpen
  \bibfield  {author} {\bibinfo {author} {\bibfnamefont {V.}~\bibnamefont
  {Khemani}}, \bibinfo {author} {\bibfnamefont {A.}~\bibnamefont {Lazarides}},
  \bibinfo {author} {\bibfnamefont {R.}~\bibnamefont {Moessner}},\ and\
  \bibinfo {author} {\bibfnamefont {S.~L.}\ \bibnamefont {Sondhi}},\ }\bibfield
   {title} {\bibinfo {title} {Phase structure of driven quantum systems},\
  }\href@noop {} {\bibfield  {journal} {\bibinfo  {journal} {Phys. Rev. Lett.}\
  }\textbf {\bibinfo {volume} {116}},\ \bibinfo {pages} {250401} (\bibinfo
  {year} {2016})}\BibitemShut {NoStop}%
\bibitem [{\citenamefont {Else}\ \emph {et~al.}(2016)\citenamefont {Else},
  \citenamefont {Bauer},\ and\ \citenamefont {Nayak}}]{Else:2016ue}%
  \BibitemOpen
  \bibfield  {author} {\bibinfo {author} {\bibfnamefont {D.~V.}\ \bibnamefont
  {Else}}, \bibinfo {author} {\bibfnamefont {B.}~\bibnamefont {Bauer}},\ and\
  \bibinfo {author} {\bibfnamefont {C.}~\bibnamefont {Nayak}},\ }\bibfield
  {title} {\bibinfo {title} {Floquet time crystals},\ }\href@noop {} {\bibfield
   {journal} {\bibinfo  {journal} {Phys. Rev. Lett.}\ }\textbf {\bibinfo
  {volume} {117}},\ \bibinfo {pages} {090402} (\bibinfo {year}
  {2016})}\BibitemShut {NoStop}%
\bibitem [{\citenamefont {Yao}\ \emph {et~al.}(2017)\citenamefont {Yao},
  \citenamefont {Potter}, \citenamefont {Potirniche},\ and\ \citenamefont
  {Vishwanath}}]{Yao2017}%
  \BibitemOpen
  \bibfield  {author} {\bibinfo {author} {\bibfnamefont {N.~Y.}\ \bibnamefont
  {Yao}}, \bibinfo {author} {\bibfnamefont {A.~C.}\ \bibnamefont {Potter}},
  \bibinfo {author} {\bibfnamefont {I.~D.}\ \bibnamefont {Potirniche}},\ and\
  \bibinfo {author} {\bibfnamefont {A.}~\bibnamefont {Vishwanath}},\ }\bibfield
   {title} {\bibinfo {title} {Discrete time crystals: Rigidity, criticality,
  and realizations},\ }\href@noop {} {\bibfield  {journal} {\bibinfo  {journal}
  {Phys. Rev. Lett.}\ }\textbf {\bibinfo {volume} {118}},\ \bibinfo {pages}
  {030401} (\bibinfo {year} {2017})}\BibitemShut {NoStop}%
\bibitem [{\citenamefont {Lan}\ \emph {et~al.}(2018)\citenamefont {Lan},
  \citenamefont {van Horssen}, \citenamefont {Powell},\ and\ \citenamefont
  {Garrahan}}]{Lan2018}%
  \BibitemOpen
  \bibfield  {author} {\bibinfo {author} {\bibfnamefont {Z.}~\bibnamefont
  {Lan}}, \bibinfo {author} {\bibfnamefont {M.}~\bibnamefont {van Horssen}},
  \bibinfo {author} {\bibfnamefont {S.}~\bibnamefont {Powell}},\ and\ \bibinfo
  {author} {\bibfnamefont {J.~P.}\ \bibnamefont {Garrahan}},\ }\bibfield
  {title} {\bibinfo {title} {Quantum slow relaxation and metastability due to
  dynamical constraints},\ }\href@noop {} {\bibfield  {journal} {\bibinfo
  {journal} {Phys. Rev. Lett.}\ }\textbf {\bibinfo {volume} {121}},\ \bibinfo
  {pages} {040603} (\bibinfo {year} {2018})}\BibitemShut {NoStop}%
\bibitem [{\citenamefont {Roy}\ and\ \citenamefont {Logan}(2019)}]{Roy2019a}%
  \BibitemOpen
  \bibfield  {author} {\bibinfo {author} {\bibfnamefont {S.}~\bibnamefont
  {Roy}}\ and\ \bibinfo {author} {\bibfnamefont {D.}~\bibnamefont {Logan}},\
  }\bibfield  {title} {\bibinfo {title} {Self-consistent theory of many-body
  localisation in a quantum spin chain with long-range interactions},\
  }\href@noop {} {\bibfield  {journal} {\bibinfo  {journal} {arxiv}\ }
  (\bibinfo {year} {2019})}\BibitemShut {NoStop}%
\bibitem [{\citenamefont {Deger}\ \emph
  {et~al.}(2022{\natexlab{a}})\citenamefont {Deger}, \citenamefont {Roy},\ and\
  \citenamefont {Lazarides}}]{Deger22a}%
  \BibitemOpen
  \bibfield  {author} {\bibinfo {author} {\bibfnamefont {A.}~\bibnamefont
  {Deger}}, \bibinfo {author} {\bibfnamefont {S.}~\bibnamefont {Roy}},\ and\
  \bibinfo {author} {\bibfnamefont {A.}~\bibnamefont {Lazarides}},\ }\bibfield
  {title} {\bibinfo {title} {Arresting {Classical} {Many}-{Body} {Chaos} by
  {Kinetic} {Constraints}},\ }\href@noop {} {\bibfield  {journal} {\bibinfo
  {journal} {Phys. Rev. Lett.}\ }\textbf {\bibinfo {volume} {129}},\ \bibinfo
  {pages} {160601} (\bibinfo {year} {2022}{\natexlab{a}})}\BibitemShut
  {NoStop}%
\bibitem [{\citenamefont {Deger}\ \emph
  {et~al.}(2022{\natexlab{b}})\citenamefont {Deger}, \citenamefont
  {Lazarides},\ and\ \citenamefont {Roy}}]{Deger22b}%
  \BibitemOpen
  \bibfield  {author} {\bibinfo {author} {\bibfnamefont {A.}~\bibnamefont
  {Deger}}, \bibinfo {author} {\bibfnamefont {A.}~\bibnamefont {Lazarides}},\
  and\ \bibinfo {author} {\bibfnamefont {S.}~\bibnamefont {Roy}},\ }\bibfield
  {title} {\bibinfo {title} {Constrained dynamics and directed percolation},\
  }\href@noop {} {\bibfield  {journal} {\bibinfo  {journal} {Phys. Rev. Lett.}\
  }\textbf {\bibinfo {volume} {129}},\ \bibinfo {pages} {190601} (\bibinfo
  {year} {2022}{\natexlab{b}})}\BibitemShut {NoStop}%
\bibitem [{\citenamefont {Basko}\ \emph {et~al.}(2006)\citenamefont {Basko},
  \citenamefont {Aleiner},\ and\ \citenamefont {Altshuler}}]{Basko:2006hh}%
  \BibitemOpen
  \bibfield  {author} {\bibinfo {author} {\bibfnamefont {D.~M.}\ \bibnamefont
  {Basko}}, \bibinfo {author} {\bibfnamefont {I.~L.}\ \bibnamefont {Aleiner}},\
  and\ \bibinfo {author} {\bibfnamefont {B.~L.}\ \bibnamefont {Altshuler}},\
  }\bibfield  {title} {\bibinfo {title} {Metaltextendashinsulator transition in
  a weakly interacting many-electron system with localized single-particle
  states},\ }\href@noop {} {\bibfield  {journal} {\bibinfo  {journal} {Ann.
  Phys. (N. Y).}\ }\textbf {\bibinfo {volume} {321}},\ \bibinfo {pages} {1126}
  (\bibinfo {year} {2006})}\BibitemShut {NoStop}%
\bibitem [{\citenamefont {Oganesyan}\ and\ \citenamefont
  {Huse}(2007)}]{Oganesyan:2007ex}%
  \BibitemOpen
  \bibfield  {author} {\bibinfo {author} {\bibfnamefont {V.}~\bibnamefont
  {Oganesyan}}\ and\ \bibinfo {author} {\bibfnamefont {D.~A.}\ \bibnamefont
  {Huse}},\ }\bibfield  {title} {\bibinfo {title} {Localization of interacting
  fermions at high temperature},\ }\href@noop {} {\bibfield  {journal}
  {\bibinfo  {journal} {Phys. Rev. B}\ }\textbf {\bibinfo {volume} {75}},\
  \bibinfo {pages} {155111} (\bibinfo {year} {2007})}\BibitemShut {NoStop}%
\bibitem [{\citenamefont {Znidaric}\ \emph {et~al.}(2008)\citenamefont
  {Znidaric}, \citenamefont {Prosen},\ and\ \citenamefont {Prelov{\v
  s}ek}}]{Znidaric:2008ux}%
  \BibitemOpen
  \bibfield  {author} {\bibinfo {author} {\bibfnamefont {M.}~\bibnamefont
  {Znidaric}}, \bibinfo {author} {\bibfnamefont {T.}~\bibnamefont {Prosen}},\
  and\ \bibinfo {author} {\bibfnamefont {P.}~\bibnamefont {Prelov{\v s}ek}},\
  }\bibfield  {title} {\bibinfo {title} {Many-body localization in the
  heisenberg xxz magnet in a random field},\ }\href@noop {} {\bibfield
  {journal} {\bibinfo  {journal} {Phys. Rev. B}\ }\textbf {\bibinfo {volume}
  {77}},\ \bibinfo {pages} {64426} (\bibinfo {year} {2008})}\BibitemShut
  {NoStop}%
\bibitem [{\citenamefont {Pal}\ and\ \citenamefont {Huse}(2010)}]{Pal:2010gr}%
  \BibitemOpen
  \bibfield  {author} {\bibinfo {author} {\bibfnamefont {A.}~\bibnamefont
  {Pal}}\ and\ \bibinfo {author} {\bibfnamefont {D.~A.}\ \bibnamefont {Huse}},\
  }\bibfield  {title} {\bibinfo {title} {Many-body localization phase
  transition},\ }\href@noop {} {\bibfield  {journal} {\bibinfo  {journal}
  {Phys. Rev. B}\ }\textbf {\bibinfo {volume} {82}},\ \bibinfo {pages} {174411}
  (\bibinfo {year} {2010})}\BibitemShut {NoStop}%
\bibitem [{\citenamefont {Anderson}(1958)}]{Anderson:1958fz}%
  \BibitemOpen
  \bibfield  {author} {\bibinfo {author} {\bibfnamefont {P.~W.}\ \bibnamefont
  {Anderson}},\ }\bibfield  {title} {\bibinfo {title} {Absence of diffusion in
  certain random lattices},\ }\href@noop {} {\bibfield  {journal} {\bibinfo
  {journal} {Phys. Rev.}\ }\textbf {\bibinfo {volume} {109}},\ \bibinfo {pages}
  {1492} (\bibinfo {year} {1958})}\BibitemShut {NoStop}%
\bibitem [{\citenamefont {Wannier}(1962)}]{wannier1962dynamics}%
  \BibitemOpen
  \bibfield  {author} {\bibinfo {author} {\bibfnamefont {G.~H.}\ \bibnamefont
  {Wannier}},\ }\bibfield  {title} {\bibinfo {title} {Dynamics of band
  electrons in electric and magnetic fields},\ }\href@noop {} {\bibfield
  {journal} {\bibinfo  {journal} {Reviews of Modern Physics}\ }\textbf
  {\bibinfo {volume} {34}},\ \bibinfo {pages} {645} (\bibinfo {year}
  {1962})}\BibitemShut {NoStop}%
\bibitem [{\citenamefont {Schulz}\ \emph {et~al.}(2019)\citenamefont {Schulz},
  \citenamefont {Hooley}, \citenamefont {Moessner},\ and\ \citenamefont
  {Pollmann}}]{schulz2019stark}%
  \BibitemOpen
  \bibfield  {author} {\bibinfo {author} {\bibfnamefont {M.}~\bibnamefont
  {Schulz}}, \bibinfo {author} {\bibfnamefont {C.}~\bibnamefont {Hooley}},
  \bibinfo {author} {\bibfnamefont {R.}~\bibnamefont {Moessner}},\ and\
  \bibinfo {author} {\bibfnamefont {F.}~\bibnamefont {Pollmann}},\ }\bibfield
  {title} {\bibinfo {title} {Stark many-body localization},\ }\href@noop {}
  {\bibfield  {journal} {\bibinfo  {journal} {Physical review letters}\
  }\textbf {\bibinfo {volume} {122}},\ \bibinfo {pages} {040606} (\bibinfo
  {year} {2019})}\BibitemShut {NoStop}%
\bibitem [{\citenamefont {Morong}\ \emph {et~al.}(2021)\citenamefont {Morong},
  \citenamefont {Liu}, \citenamefont {Becker}, \citenamefont {Collins},
  \citenamefont {Feng}, \citenamefont {Kyprianidis}, \citenamefont {Pagano},
  \citenamefont {You}, \citenamefont {Gorshkov},\ and\ \citenamefont
  {Monroe}}]{Morong2021}%
  \BibitemOpen
  \bibfield  {author} {\bibinfo {author} {\bibfnamefont {W.}~\bibnamefont
  {Morong}}, \bibinfo {author} {\bibfnamefont {F.}~\bibnamefont {Liu}},
  \bibinfo {author} {\bibfnamefont {P.}~\bibnamefont {Becker}}, \bibinfo
  {author} {\bibfnamefont {K.}~\bibnamefont {Collins}}, \bibinfo {author}
  {\bibfnamefont {L.}~\bibnamefont {Feng}}, \bibinfo {author} {\bibfnamefont
  {A.}~\bibnamefont {Kyprianidis}}, \bibinfo {author} {\bibfnamefont
  {G.}~\bibnamefont {Pagano}}, \bibinfo {author} {\bibfnamefont
  {T.}~\bibnamefont {You}}, \bibinfo {author} {\bibfnamefont {A.}~\bibnamefont
  {Gorshkov}},\ and\ \bibinfo {author} {\bibfnamefont {C.}~\bibnamefont
  {Monroe}},\ }\bibfield  {title} {\bibinfo {title} {Observation of stark
  many-body localization without disorder.},\ }\href@noop {} {\bibfield
  {journal} {\bibinfo  {journal} {Nature}\ }\textbf {\bibinfo {volume} {599}},\
  \bibinfo {pages} {393} (\bibinfo {year} {2021})}\BibitemShut {NoStop}%
\bibitem [{\citenamefont {Taylor}\ \emph {et~al.}(2020)\citenamefont {Taylor},
  \citenamefont {Schulz}, \citenamefont {Pollmann},\ and\ \citenamefont
  {Moessner}}]{taylor2020experimental}%
  \BibitemOpen
  \bibfield  {author} {\bibinfo {author} {\bibfnamefont {S.~R.}\ \bibnamefont
  {Taylor}}, \bibinfo {author} {\bibfnamefont {M.}~\bibnamefont {Schulz}},
  \bibinfo {author} {\bibfnamefont {F.}~\bibnamefont {Pollmann}},\ and\
  \bibinfo {author} {\bibfnamefont {R.}~\bibnamefont {Moessner}},\ }\bibfield
  {title} {\bibinfo {title} {Experimental probes of stark many-body
  localization},\ }\href@noop {} {\bibfield  {journal} {\bibinfo  {journal}
  {Physical Review B}\ }\textbf {\bibinfo {volume} {102}},\ \bibinfo {pages}
  {054206} (\bibinfo {year} {2020})}\BibitemShut {NoStop}%
\bibitem [{\citenamefont {Guo}\ \emph {et~al.}(2021)\citenamefont {Guo},
  \citenamefont {Cheng}, \citenamefont {Li}, \citenamefont {Xu}, \citenamefont
  {Zhang}, \citenamefont {Wang}, \citenamefont {Song}, \citenamefont {Liu},
  \citenamefont {Ren}, \citenamefont {Dong} \emph {et~al.}}]{guo2021stark}%
  \BibitemOpen
  \bibfield  {author} {\bibinfo {author} {\bibfnamefont {Q.}~\bibnamefont
  {Guo}}, \bibinfo {author} {\bibfnamefont {C.}~\bibnamefont {Cheng}}, \bibinfo
  {author} {\bibfnamefont {H.}~\bibnamefont {Li}}, \bibinfo {author}
  {\bibfnamefont {S.}~\bibnamefont {Xu}}, \bibinfo {author} {\bibfnamefont
  {P.}~\bibnamefont {Zhang}}, \bibinfo {author} {\bibfnamefont
  {Z.}~\bibnamefont {Wang}}, \bibinfo {author} {\bibfnamefont {C.}~\bibnamefont
  {Song}}, \bibinfo {author} {\bibfnamefont {W.}~\bibnamefont {Liu}}, \bibinfo
  {author} {\bibfnamefont {W.}~\bibnamefont {Ren}}, \bibinfo {author}
  {\bibfnamefont {H.}~\bibnamefont {Dong}}, \emph {et~al.},\ }\bibfield
  {title} {\bibinfo {title} {Stark many-body localization on a superconducting
  quantum processor},\ }\href@noop {} {\bibfield  {journal} {\bibinfo
  {journal} {Physical review letters}\ }\textbf {\bibinfo {volume} {127}},\
  \bibinfo {pages} {240502} (\bibinfo {year} {2021})}\BibitemShut {NoStop}%
\bibitem [{\citenamefont {Zhang}\ \emph {et~al.}(2021)\citenamefont {Zhang},
  \citenamefont {Ke}, \citenamefont {Liu},\ and\ \citenamefont
  {Lee}}]{zhang2021}%
  \BibitemOpen
  \bibfield  {author} {\bibinfo {author} {\bibfnamefont {L.}~\bibnamefont
  {Zhang}}, \bibinfo {author} {\bibfnamefont {Y.}~\bibnamefont {Ke}}, \bibinfo
  {author} {\bibfnamefont {W.}~\bibnamefont {Liu}},\ and\ \bibinfo {author}
  {\bibfnamefont {C.}~\bibnamefont {Lee}},\ }\bibfield  {title} {\bibinfo
  {title} {Mobility edge of stark many-body localization},\ }\href@noop {}
  {\bibfield  {journal} {\bibinfo  {journal} {Physical Review A}\ }\textbf
  {\bibinfo {volume} {103}} (\bibinfo {year} {2021})}\BibitemShut {NoStop}%
\bibitem [{\citenamefont {Doggen}\ \emph {et~al.}(2021)\citenamefont {Doggen},
  \citenamefont {Gornyi},\ and\ \citenamefont {Polyakov}}]{doggen2021stark}%
  \BibitemOpen
  \bibfield  {author} {\bibinfo {author} {\bibfnamefont {E.~V.}\ \bibnamefont
  {Doggen}}, \bibinfo {author} {\bibfnamefont {I.~V.}\ \bibnamefont {Gornyi}},\
  and\ \bibinfo {author} {\bibfnamefont {D.~G.}\ \bibnamefont {Polyakov}},\
  }\bibfield  {title} {\bibinfo {title} {Stark many-body localization: Evidence
  for hilbert-space shattering},\ }\href@noop {} {\bibfield  {journal}
  {\bibinfo  {journal} {Physical Review B}\ }\textbf {\bibinfo {volume}
  {103}},\ \bibinfo {pages} {L100202} (\bibinfo {year} {2021})}\BibitemShut
  {NoStop}%
\bibitem [{\citenamefont {Holthaus}\ and\ \citenamefont
  {Hone}(1996)}]{Holthaus1996}%
  \BibitemOpen
  \bibfield  {author} {\bibinfo {author} {\bibfnamefont {M.}~\bibnamefont
  {Holthaus}}\ and\ \bibinfo {author} {\bibfnamefont {D.~W.}\ \bibnamefont
  {Hone}},\ }\href@noop {} {\emph {\bibinfo {title} {Localization Effects in
  ac-driven Tight-Binding Lattices}}},\ Vol.\ \bibinfo {volume} {2812(March)}\
  (\bibinfo  {publisher} {Cornell University Library},\ \bibinfo {year}
  {1996})\ pp.\ \bibinfo {pages} {105--137}\BibitemShut {NoStop}%
\bibitem [{\citenamefont {Shon}\ and\ \citenamefont
  {Nazareno}(1992)}]{Shon1992}%
  \BibitemOpen
  \bibfield  {author} {\bibinfo {author} {\bibfnamefont {N.~H.}\ \bibnamefont
  {Shon}}\ and\ \bibinfo {author} {\bibfnamefont {H.}~\bibnamefont
  {Nazareno}},\ }\bibfield  {title} {\bibinfo {title} {On the dynamic
  localization in 1d tight-binding systems},\ }\href@noop {} {\bibfield
  {journal} {\bibinfo  {journal} {Journal of Physics: Condensed Matter}\
  }\textbf {\bibinfo {volume} {4}},\ \bibinfo {pages} {L611} (\bibinfo {year}
  {1992})}\BibitemShut {NoStop}%
\bibitem [{\citenamefont {Ponte}\ \emph {et~al.}(2015)\citenamefont {Ponte},
  \citenamefont {Papi{\'c}}, \citenamefont {Huveneers},\ and\ \citenamefont
  {Abanin}}]{Ponte:2015dc}%
  \BibitemOpen
  \bibfield  {author} {\bibinfo {author} {\bibfnamefont {P.}~\bibnamefont
  {Ponte}}, \bibinfo {author} {\bibfnamefont {Z.}~\bibnamefont {Papi{\'c}}},
  \bibinfo {author} {\bibfnamefont {F.}~\bibnamefont {Huveneers}},\ and\
  \bibinfo {author} {\bibfnamefont {D.~A.}\ \bibnamefont {Abanin}},\ }\bibfield
   {title} {\bibinfo {title} {Many-body localization in periodically driven
  systems},\ }\href@noop {} {\bibfield  {journal} {\bibinfo  {journal} {Phys.
  Rev. Lett.}\ }\textbf {\bibinfo {volume} {114}},\ \bibinfo {pages} {140401}
  (\bibinfo {year} {2015})}\BibitemShut {NoStop}%
\bibitem [{\citenamefont {Bhakuni}\ \emph {et~al.}(2020)\citenamefont
  {Bhakuni}, \citenamefont {Nehra},\ and\ \citenamefont
  {Sharma}}]{Bhakuni2020}%
  \BibitemOpen
  \bibfield  {author} {\bibinfo {author} {\bibfnamefont {D.~S.}\ \bibnamefont
  {Bhakuni}}, \bibinfo {author} {\bibfnamefont {R.}~\bibnamefont {Nehra}},\
  and\ \bibinfo {author} {\bibfnamefont {A.}~\bibnamefont {Sharma}},\
  }\bibfield  {title} {\bibinfo {title} {Drive-induced many-body localization
  and coherent destruction of stark many-body localization},\ }\href@noop {}
  {\bibfield  {journal} {\bibinfo  {journal} {Physical Review B}\ }\textbf
  {\bibinfo {volume} {102}} (\bibinfo {year} {2020})}\BibitemShut {NoStop}%
\bibitem [{\citenamefont {Kloss}\ \emph {et~al.}(2023)\citenamefont {Kloss},
  \citenamefont {Halimeh}, \citenamefont {Lazarides},\ and\ \citenamefont
  {Bar~Lev}}]{kloss_absence_2023}%
  \BibitemOpen
  \bibfield  {author} {\bibinfo {author} {\bibfnamefont {B.}~\bibnamefont
  {Kloss}}, \bibinfo {author} {\bibfnamefont {J.~C.}\ \bibnamefont {Halimeh}},
  \bibinfo {author} {\bibfnamefont {A.}~\bibnamefont {Lazarides}},\ and\
  \bibinfo {author} {\bibfnamefont {Y.}~\bibnamefont {Bar~Lev}},\ }\bibfield
  {title} {\bibinfo {title} {Absence of localization in interacting spin chains
  with a discrete symmetry},\ }\href@noop {} {\bibfield  {journal} {\bibinfo
  {journal} {Nature Communications}\ }\textbf {\bibinfo {volume} {14}},\
  \bibinfo {pages} {3778} (\bibinfo {year} {2023})}\BibitemShut {NoStop}%
\bibitem [{\citenamefont {Shirley}(1965)}]{Shirley:1965cy}%
  \BibitemOpen
  \bibfield  {author} {\bibinfo {author} {\bibfnamefont {J.~H.}\ \bibnamefont
  {Shirley}},\ }\bibfield  {title} {\bibinfo {title} {Solution of the
  schr{\"o}dinger equation with a hamiltonian periodic in time},\ }\href@noop
  {} {\bibfield  {journal} {\bibinfo  {journal} {Phys. Rev.}\ }\textbf
  {\bibinfo {volume} {138}},\ \bibinfo {pages} {B979} (\bibinfo {year}
  {1965})}\BibitemShut {NoStop}%
\bibitem [{\citenamefont {Lazarides}\ \emph {et~al.}(2014)\citenamefont
  {Lazarides}, \citenamefont {Das},\ and\ \citenamefont
  {Moessner}}]{Lazarides:2014ie}%
  \BibitemOpen
  \bibfield  {author} {\bibinfo {author} {\bibfnamefont {A.}~\bibnamefont
  {Lazarides}}, \bibinfo {author} {\bibfnamefont {A.}~\bibnamefont {Das}},\
  and\ \bibinfo {author} {\bibfnamefont {R.}~\bibnamefont {Moessner}},\
  }\bibfield  {title} {\bibinfo {title} {Equilibrium states of generic quantum
  systems subject to periodic driving},\ }\href@noop {} {\bibfield  {journal}
  {\bibinfo  {journal} {Phys. Rev. E}\ }\textbf {\bibinfo {volume} {90}},\
  \bibinfo {pages} {12110} (\bibinfo {year} {2014})}\BibitemShut {NoStop}%
\bibitem [{\citenamefont {De~Luca}\ and\ \citenamefont
  {Scardicchio}(2013)}]{DeLuca2013}%
  \BibitemOpen
  \bibfield  {author} {\bibinfo {author} {\bibfnamefont {A.}~\bibnamefont
  {De~Luca}}\ and\ \bibinfo {author} {\bibfnamefont {A.}~\bibnamefont
  {Scardicchio}},\ }\bibfield  {title} {\bibinfo {title} {Ergodicity breaking
  in a model showing many-body localization},\ }\href@noop {} {\bibfield
  {journal} {\bibinfo  {journal} {Europhysics Letters}\ }\textbf {\bibinfo
  {volume} {101}},\ \bibinfo {pages} {37003} (\bibinfo {year}
  {2013})}\BibitemShut {NoStop}%
\bibitem [{\citenamefont {Luitz}\ \emph {et~al.}(2020)\citenamefont {Luitz},
  \citenamefont {Khaymovich},\ and\ \citenamefont {Bar~Lev}}]{Luitz2020}%
  \BibitemOpen
  \bibfield  {author} {\bibinfo {author} {\bibfnamefont {D.~J.}\ \bibnamefont
  {Luitz}}, \bibinfo {author} {\bibfnamefont {I.}~\bibnamefont {Khaymovich}},\
  and\ \bibinfo {author} {\bibfnamefont {Y.}~\bibnamefont {Bar~Lev}},\
  }\bibfield  {title} {\bibinfo {title} {Multifractality and its role in
  anomalous transport in the disordered xxz spin-chain},\ }\href@noop {}
  {\bibfield  {journal} {\bibinfo  {journal} {SciPost Physics Core}\ }\textbf
  {\bibinfo {volume} {2}} (\bibinfo {year} {2020})}\BibitemShut {NoStop}%
\bibitem [{\citenamefont {Luitz}\ \emph {et~al.}(2015)\citenamefont {Luitz},
  \citenamefont {Laflorencie},\ and\ \citenamefont {Alet}}]{Luitz:2015iv}%
  \BibitemOpen
  \bibfield  {author} {\bibinfo {author} {\bibfnamefont {D.~J.}\ \bibnamefont
  {Luitz}}, \bibinfo {author} {\bibfnamefont {N.}~\bibnamefont {Laflorencie}},\
  and\ \bibinfo {author} {\bibfnamefont {F.}~\bibnamefont {Alet}},\ }\bibfield
  {title} {\bibinfo {title} {Many-body localization edge in the random-field
  heisenberg chain},\ }\href@noop {} {\bibfield  {journal} {\bibinfo  {journal}
  {Phys. Rev. B}\ }\textbf {\bibinfo {volume} {91}},\ \bibinfo {pages} {81103}
  (\bibinfo {year} {2015})}\BibitemShut {NoStop}%
\bibitem [{\citenamefont {Serbyn}\ \emph {et~al.}(2017)\citenamefont {Serbyn},
  \citenamefont {Papi{\'c}},\ and\ \citenamefont
  {Abanin}}]{serbyn2017thouless}%
  \BibitemOpen
  \bibfield  {author} {\bibinfo {author} {\bibfnamefont {M.}~\bibnamefont
  {Serbyn}}, \bibinfo {author} {\bibfnamefont {Z.}~\bibnamefont {Papi{\'c}}},\
  and\ \bibinfo {author} {\bibfnamefont {D.~A.}\ \bibnamefont {Abanin}},\
  }\bibfield  {title} {\bibinfo {title} {Thouless energy and multifractality
  across the many-body localization transition},\ }\href@noop {} {\bibfield
  {journal} {\bibinfo  {journal} {Physical Review B}\ }\textbf {\bibinfo
  {volume} {96}},\ \bibinfo {pages} {104201} (\bibinfo {year}
  {2017})}\BibitemShut {NoStop}%
\bibitem [{\citenamefont {Bar~Lev}\ and\ \citenamefont
  {Lazarides}(2024)}]{bar_lev_discrete_2024}%
  \BibitemOpen
  \bibfield  {author} {\bibinfo {author} {\bibfnamefont {Y.}~\bibnamefont
  {Bar~Lev}}\ and\ \bibinfo {author} {\bibfnamefont {A.}~\bibnamefont
  {Lazarides}},\ }\href@noop {} {\bibinfo {title} {Discrete time-crystals in
  linear potentials}} (\bibinfo {year} {2024}),\ \bibinfo {note}
  {arXiv:2403.01912 [cond-mat]}\BibitemShut {NoStop}%
\end{thebibliography}

%

\newpage 
\appendix

\section{Scanning Parameters}

One question is how the behaviours seen depend on other parameters, such as \(J/V\). In particular, we are interested in understanding whether there is an optimum value of \(J/V\) for which the distinction between MBL and ETH becomes most apparent for a given driving amplitude. One way to explore this is to calculate the standard deviation of the EEVs both on resonance and off resonance, and calculate their difference as a function of \(J/V\). This is demonstrated in Fig.~\ref{fig/UC_2}(a), with the quantity
\begin{equation}\label{delta_sigma}
\Delta\sigma=|\sigma(n=2.0)-\sigma(n=0.7)|,
\end{equation}
where \(\sigma\) is the standard deviation of the EEV probability distribution \(P(\langle n_6\rangle)\). Here \(J\) is the quantity being changed, while \(V=0.5\). The resulting distribution demonstrates the interplay between the limits of trivial localisation \(J=0\) and delocalisation \(J>>V,\gamma\), where the peak indicates the point of clearest distinction between resonant and off-resonant driving.

\begin{figure}[H]
    \centering
    \includegraphics[width=1.0\linewidth]{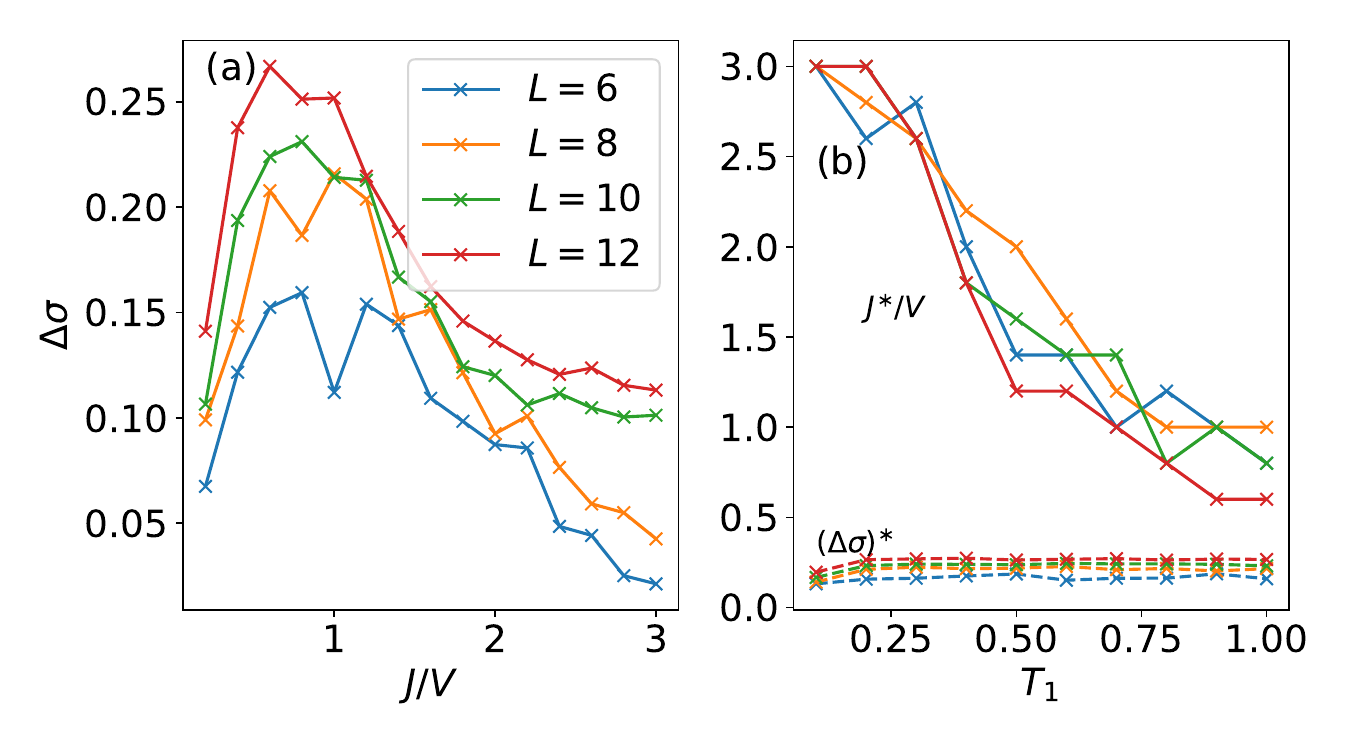}
    \caption{(a) Difference in the standard deviation of \(P(\langle n_6\rangle)\) for resonant (\(n=2.0\)) and off-resonant (\(n=0.7\)) driving as defined in equation \ref{delta_sigma}, plotted as a function of \(J/V\). Remaining parameters are \(V=0.5,\,\gamma=5,\,T_1=1.0\). (b) Maximum \(\Delta\sigma\) and corresponding \(J/V\), denoted as \((\Delta\sigma)^{\ast}\) and \(J^{\ast}/V\), respectively, plotted as a function of \(T_1\).}
    \label{fig/delta_sigma}
\end{figure}

Next, we obtain the quantities \((\Delta\sigma)^{\ast}\) and \(J^{\ast}/V\), which are defined as the maximum value of \(\Delta\sigma\) and the value of \(J/V\) at which this exists, respectively. These quantities are then scanned against \(T_1\), as shown in Fig.~\ref{fig/delta_sigma}(b). \(J^{\ast}/V\) can be roughly seen to decrease with \(T_1\). This is unsurprising, as both \(T_1\) and \(J\) delocalise the system, meaning that at larger \(T_1\), lower \(J\) is needed in order for MBL to persist off resonance. Importantly, we can see that \((\Delta\sigma)^{\ast}\) is independent of \(T_1\), suggesting that for any amplitude of driving, we will see a similar distinction provided \(J/V=J^{\ast}/V\). Finally, we note that the values \(J=V=0.5\) correspond approximately to \(J^{\ast}/V\) at \(T_1=0.5\), which are the parameters used in Fig.~\ref{fig/eev_hist}, suggesting that the distinction demonstrated is optimum.

\section{Quasienergy Structure}

In the left column in Fig.~\ref{fig/eev_hist}, we can see that the density of states exhibits a weak dependence on the quasienergy while off resonance, and a strong dependence while on resonance. One question of interest is whether this bandstructure will have a significant impact on the structure of the EEVs.

\begin{figure}[t]
    \centering
    \includegraphics[width=1.0\linewidth]{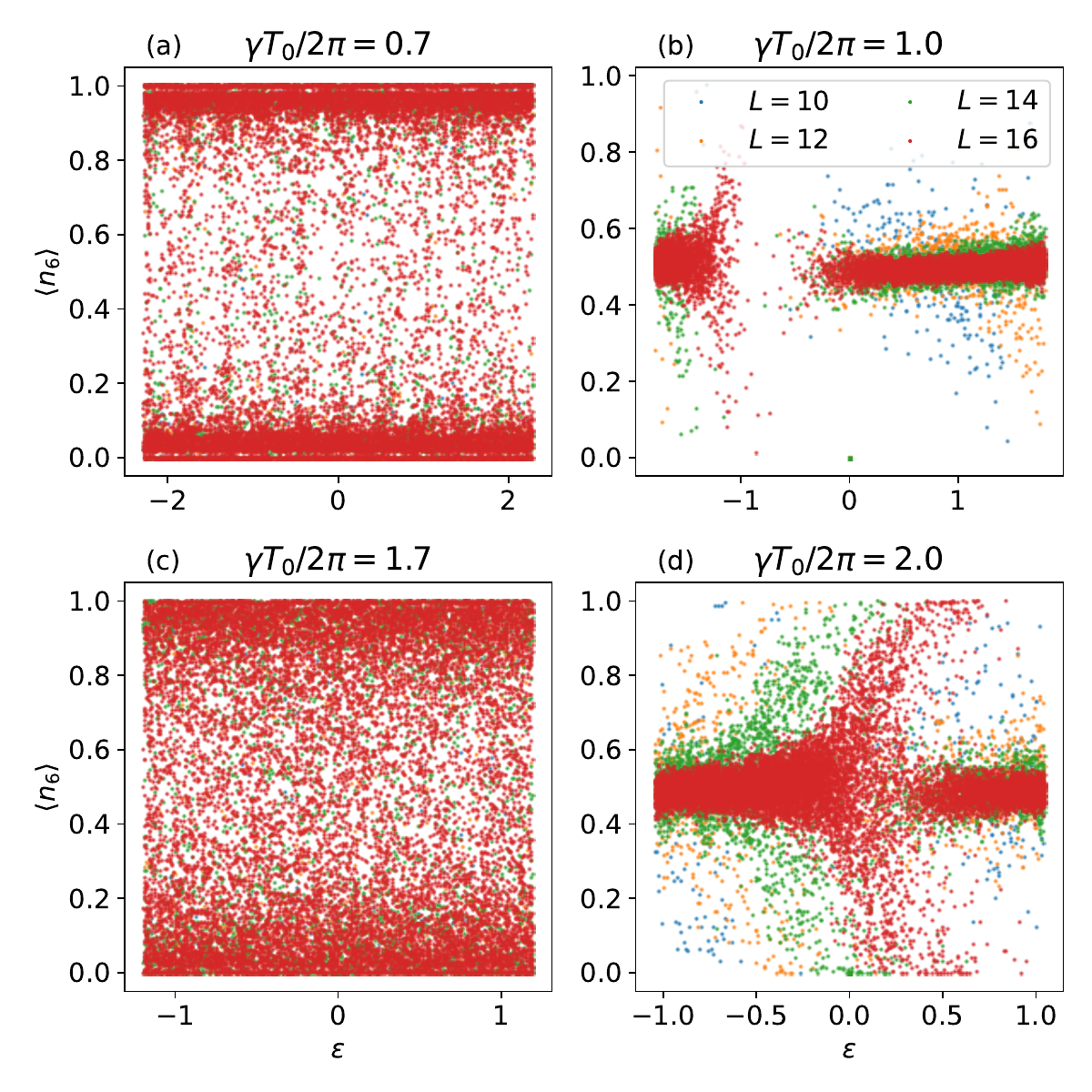}
    \caption{Scatter plot of the EEVs \(\langle n_6\rangle\) corresponding to the probability distributions shown in the right column of figure \ref{fig/eev_hist}. The left column shows cases of off-resonant driving, while the right columns correspond to resonant driving. Violation of the ETH is clearly demonstrated off resonance, with the strong fluctuation of EEVs. Meanwhile, the resonant plots showcase agreement with the ETH, except at specific regions within the quasienergy spectrum, where the EEVs are significantly more scattered.}
    \label{fig/eev_scatter}
\end{figure}

In Fig.~\ref{fig/eev_scatter}, it can be seen that during resonant driving, the EEVs obey the eigenstate thermalisation hypothesis except at specific regions within the quasienergy spectrum. We propose that this effect is due to massive degeneracies induced by resonant driving in the \(T_1=w=0\) limit, creating a clear bandstructure in the quasienergies of the Floquet operator \(\mathcal{F}=e^{-iH_0T_0}\). Let us look at the case of the second resonant point, \(n=2.0\). Initially, if we also set \(V=0\), the eigenvalues of \(H_0\) will have the form \(E=k\gamma\), where \(k\) is an integer. The corresponding Floquet eigenvalues are then \(\epsilon=e^{-4\pi ik}\)=1. Introducing nearest-neighbour interactions, we have \(E=k\gamma+aV\), with \(a\in\{0,L/2-1\}\) denoting the total number of interactions for a given Fock state. The Floquet eigenvalues are now \(\epsilon=e^{-4\pi i(k+aV/\gamma)}=e^{-4\pi iaV/\gamma}\). Here the eigenvalues split into 5 different degeneracies for sizes \(L\ge 12\), i.e. above \(L=12\) no further degeneracies are added (for our parameter values). The size of each of these degeneracies is determined by how many Fock states possess a given total nearest-neighbour interaction. For a total interaction number \(a\), this is determined by the product of binomial coefficients
\begin{equation}
m(a)=\binom{L/2+1}{a+1}\binom{L/2-1}{a}.
\end{equation}

\begin{figure}[t]
    \centering
    \includegraphics[width=1\linewidth]{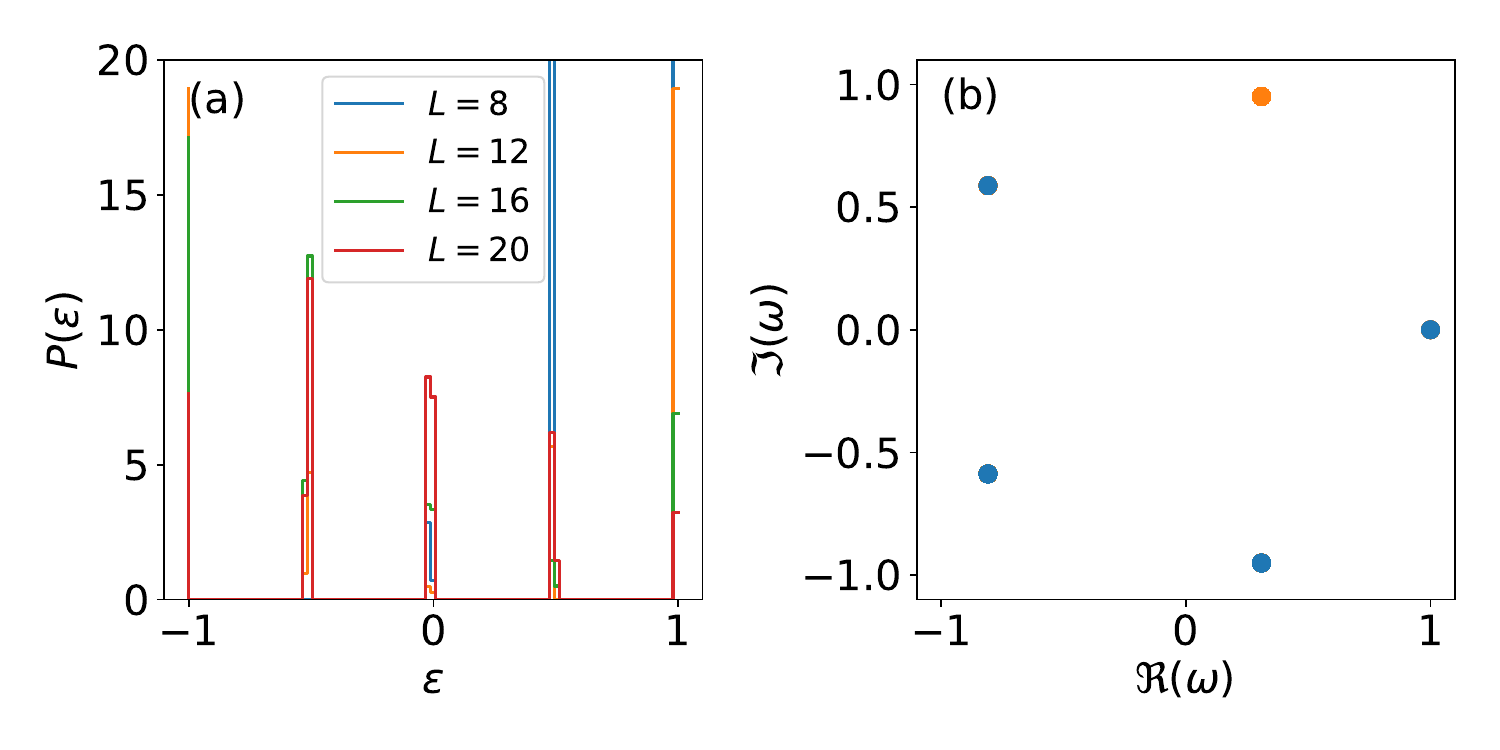}
    \caption{(a) Probability distribution \(P(\epsilon)\) of the quasienergies of the Floquet operator \(\mathcal{F}\), in the \(T_1=w=0\) limit at the second resonant point \(n=2.0\). Other parameters used are \(V=0.5,\,\gamma=5\). (b) Distribution of the eigenvalues of the Floquet operator corresponding around the unit circle. The axes correspond to the real and imaginary parts. The eigenvalues are equidistant around the unit circle, resulting in the peaked distribution of quasienergies shown in figure (a). This representation shows why no more degeneracies appear above size \(L=12\).}
    \label{fig/UC_1}
\end{figure}

\begin{figure}[t]
    \centering
    \includegraphics[width=1.0\linewidth]{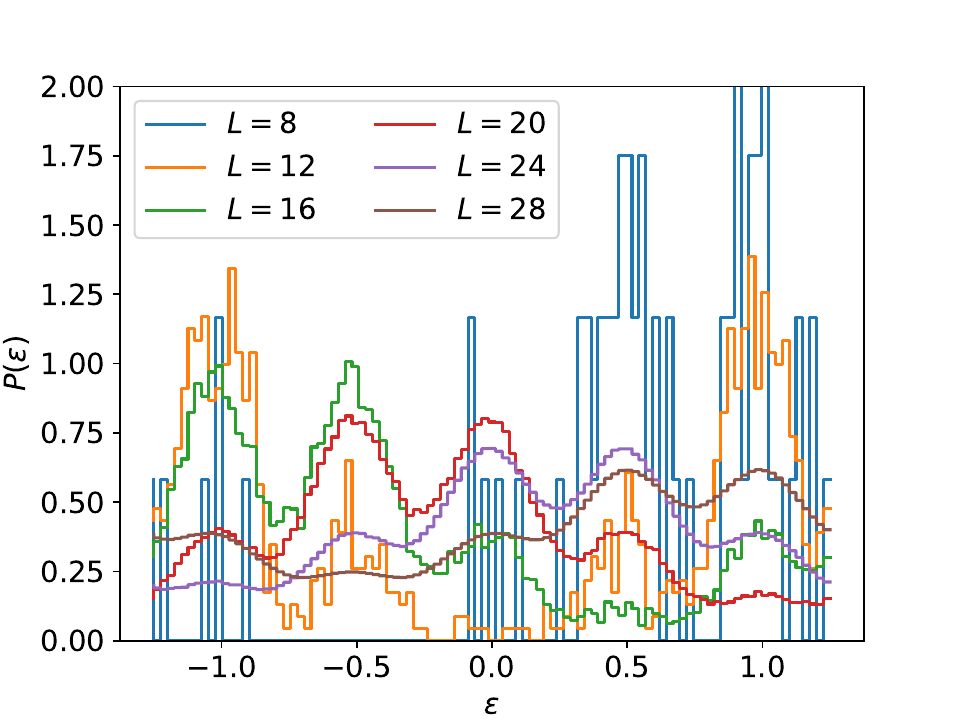}
    \caption{(a) Probability distribution of quasienergies of \(F\) in the \(T_1=0\) limit, at the second resonant point \(n=2.0\). Parameters used are \(V=0.5,\,\gamma=5,\,w=0.05\).}
    \label{fig/UC_2}
\end{figure}

The distribution of the Floquet eigenvalues around the complex unit circle is shown in Fig.~\ref{fig/UC_1}(b), along with the probability distribution of the corresponding quasienergies in (a). Next, if we re-introduce small disorder \(w\), these degeneracies will split into bands. In Fig.~\ref{fig/UC_2} we plot \(P(\epsilon)\) by calculating the eigenvalues analytically, which reveals how the dips in the distributions correspond to the scattering in the EEVs shown in Fig.~\ref{fig/eev_scatter}(d) (in particular, the dips are shown to shift right with increasing \(L\), which can be seen to match the shift of the EEVs).)

This behaviour can now be understood as follows: resonant driving eliminates energy differences introduced by the tilted potential term in the eigenvalues of \(\mathcal{F}\), resulting in a bandstructure that is governed by the distribution of nearest-neighbour interactions in \(H_0\). This creates specific regions within the quasienergy spectrum that are sparsely populated, ensuring that when driving is introduced the eigenstates in this region are subject to less mixing with the rest of the eigenstates. As a result, the eigenstates outside of these regions become similar to each other in accordance with Floquet ETH, while the eigenstates inside the region remain relatively unchanged.

The question of physical interest now is whether this behaviour persists in the thermodynamic limit, or if it is a finite-size effect. We can see from Fig.~\ref{fig/UC_2} that \(P(\epsilon)\) becomes more evenly distributed with increasing size, suggesting that the weakly mixed regions vanish at infinite size. This is also implied by the reduction in scattering with increasing size seen in the EEVs in Fig.~\ref{fig/eev_scatter}.

\end{document}